# Understanding Mechanical Properties and Failure Mechanism of Germanium-Silicon Alloy at Nanoscale

Md. Habibur Rahman[a], Emdadul Haque Chowdhury[a], Md Mahbubul Islam*[b]

[a]Department of Mechanical Engineering, Bangladesh University of Engineering and Technology, Dhaka-1000, Bangladesh

[b]Department of Mechanical Engineering, Wayne State University, Detroit MI – 48202, United States

*Corresponding Author. E-mail address: mahbub.islam@wayne.edu

## Abstract

We used molecular dynamics (MD) simulations to investigate the mechanical properties of cubic zinc blende (ZB) $Si_{0.5}Ge_{0.5}$ alloy nanowire (NW). Tersoff potential is employed to elucidate the effect of nanowire size, crystal orientations, and temperature on the material properties. We found that the reduction in the cross-sectional area results in lower ultimate tensile strength and Young's modulus of this alloy which can be attributed to the increased surface to volume ratio. The [111] oriented $Si_{0.5}Ge_{0.5}$ NW exhibits the highest fracture strength compared to other crystal orientations but [110] orientation possesses the highest fracture toughness. The effect of temperature depicts an inverse relationship with the ultimate tensile strength and Young's modulus. The increased temperature facilitates the failure of the material, thus degrades the materials' strength. Our study reveals that the vacancy defects introduced via removal of either Si or Ge atoms exhibit similar behavior, and with the increase in vacancy concentration, both ultimate tensile strength and Young's modulus reduces linearly. We further illustrate the failure characteristics of $Si_{0.5}Ge_{0.5}$ NW at two extremely low and high temperatures. The intrinsic failure characteristics of $Si_{0.5}Ge_{0.5}$ alloy is found to be insensitive to the temperature. Interestingly, at both temperatures, with the increasing strain, the cross-section of $Si_{0.5}Ge_{0.5}$ eventually resembles a neck as typically observed in ductile materials, although the NW failure is brittle in nature. Overall, this work offers a new perspective on understanding material properties and failure characteristics of ZB $Si_{0.5}Ge_{0.5}$ NW that will be a guide for designing Si-Ge based nanodevices.

## 1. Introduction

Silicon-Germanium ($Si_xGe_{1-x}$) is an alloy with any molar ratio of Silicon (Si) and Germanium (Ge). $Si_xGe_{1-x}$ is a semiconducting and thermoelectric material that has been successfully deployed in heterojunction bipolar transistor, strained metal oxide semiconductor (MOS), complementary metal oxide semiconductor (CMOS), and many other electronic systems [1], [2]. Recently, nanowires (NW) have received significant attention in the scientific community for their distinctive thermal, mechanical, electronic, and optical properties and extensively studied for potential applications as building blocks in nanoelectromechanical devices (NEMS) [3]–[5]. The NWs are excellent candidates for optoelectronic and nanoelectronics devices [6], detectors, and sensors for biological and chemical applications [7], ultrahigh-frequency resonators [8], and energy harvesting [9]. The Si NWs are extensively studied because of their exceptional properties and enormous applications in electrical and optical nanodevices and NEMS [10]–[13]. The mechanical deformation behavior and fracture mechanisms of Si NWs are probed using various experimental techniques such as bending in an atomic force microscope (AFM), tensile tests in a scanning electron microscope (SEM), transmission electron microscope (TEM) along with atomistic scale simulations [14]–[19]. Lee and Rudd [20] explored the mechanical properties of hydrogenated Si nanowire employing first-principles calculations. Likewise, investigations in Ge NWs revealed interesting properties. Smith *et al*. [21] found that Young's modulus remains independent of change in NW diameter, while Wu *et al*. [22] reported extraordinary mechanical strength for Ge nanowire and thus ability to store elastic potential energy. Mingo *et al*. investigated the mechanical and thermal properties of the Si and Ge nanostructures [23].

Silicon-germanium alloys are of great interest for many applications, such as high-mobility transistors and thermoelectric devices [24], [25]. The Si–Ge alloys are a perfect model system to analyze the possibilities and limitations of atomic-scale simulation. Ge and Si both have a ground-state diamond-lattice structure [26], [27]. Several theoretical studies have been conducted to perform local structural analysis of $Si_xGe_{1-x}$ alloys at an empirical or at the semi-empirical level. The structures have been found experimentally and with simulation [28] as truly random, with no considerable long- or short-range compositional ordering. The thermal resistivity of Si-Ge alloy was inspected with molecular dynamics (MD) simulation by Skye and Schelling [29]. A combined electronic-structure and statistical-mechanical approach were adopted to disclose the thermodynamic properties of $Si_xGe_{1-x}$ alloys by Qteish and Resta [30]. Georgakaki *et al*. [31] performed MD simulations to investigate the vibrational and mechanical properties of clamped-clamped rectangular $Si_xGe_{1-x}$ and $Si/Si_xGe_{1-x}$ nanowires (NWs). They investigated the beat vibration phenomenon, the frequency response, and determined the mechanical properties such as quality factor and Young's modulus. A contemporary work by Ma *et al*. [32] reported the tuning of Si/Ge alloy superlattices mechanical properties by changing the composition of Ge. Kim *et al*. [33] examined the size effect on Young's modulus of Si/Ge alloy using MD simulations. Amato *et al*. [34] recently presented an analysis covering a wide range of applications of Si/Ge nanowires. Furthermore, lattice defects are inevitable in nanomaterials during growth and processing; moreover, physical methods such as stress, sublimation, and irradiation can also lead to a substantial concentration of such defects [35]. Vacancy defects are lattice sites which in a perfect crystal would be occupied, but instead remain vacant. Along with affecting mechanical properties significantly [36], they also cause localization of lattice vibration around the defects.

Despite the efforts put forth to understand the properties of the Si/Ge alloys, the issues such as the effect of temperature, cross-sectional area, various defects and crystal orientation on mechanical properties and failure mechanisms remain elusive. To the best of our knowledge, although there are several studies for the inspection of Si/Ge nanowire as potential thermoelectric [9] or electronic devices [32], a comprehensive study for the mechanical properties of $Si_{0.5}Ge_{0.5}$ alloy and its failure mechanism has not been reported yet. Therefore, the current work aims to investigate the mechanical properties such as ultimate tensile strength (UTS), Young's modulus of the $Si_{0.5}Ge_{0.5}$ alloy by changing various parameters such as temperature, cross-sectional area, loading direction, and the concentration of vacancy defects using MD simulations. To understand the effect of vacancy defect on $Si_{0.5}Ge_{0.5}$ alloy, vacancies are created by removing Si and Ge atoms randomly from the alloy matrix and corresponding mechanical properties are recorded for in-depth analysis. Finally, the failure mechanism of $Si_{0.5}Ge_{0.5}$ at 100 K and 600 K temperature has been elucidated as well.

## 2. Computational Method

We used molecular dynamics (MD) simulations using the Large-scale Atomic/Molecular Massively Parallel Simulator (LAMMPS) package for calculating the tensile properties of $Si_{0.5}Ge_{0.5}$ alloy [37]. We employed Tersoff potential to describe the interatomic interactions between Ge-Ge, Si-Si, and Si-Ge atoms in the MD simulations [38]. OVITO is used for all the atomistic visualizations [39].

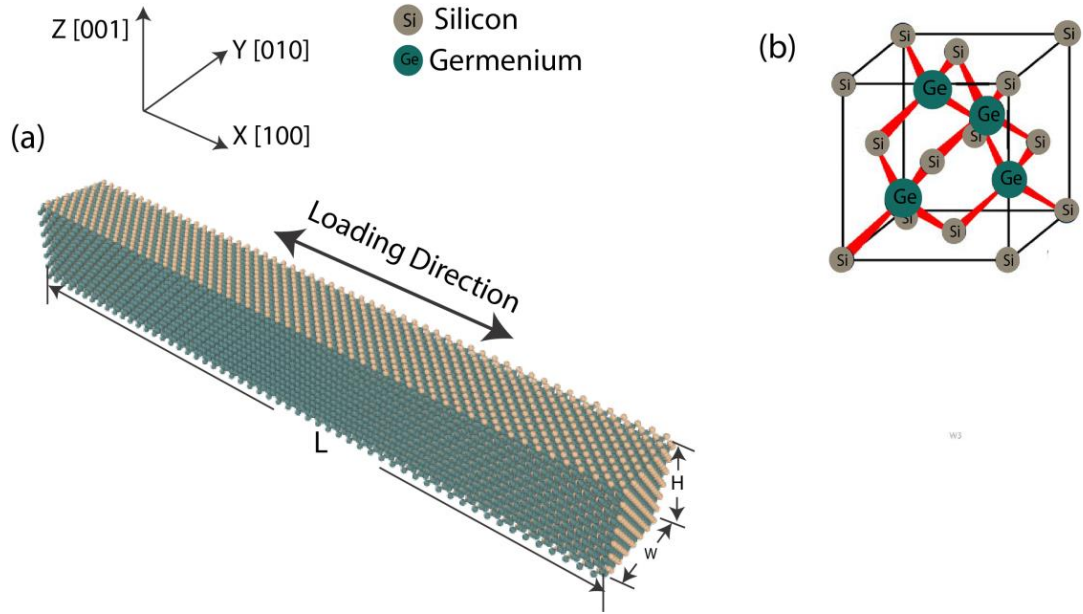

Figure 1. (a) Initial coordinates of 31.60 nm x 3.92 nm x 3.92 nm $Si_{0.5}Ge_{0.5}$. (b) Crystal structure of cubic zinc blende $Si_{0.5}Ge_{0.5}$ alloy having a lattice constant of 0.554 nm [38]

To investigate the effect of surface on the physical properties of the alloy, the cross-sectional area of the nanostructure is varied from 11.05 nm² to 19.65 nm² to maintain a length to width ratio of 8:1 [40], [41]. Three different crystal orientations are chosen, such as [100], [110], and [111] to examine the directional dependency on the material properties of the alloy. Periodic boundary conditions are applied along the loading direction (X), and other directions are kept free [40]. The equations of motion were integrated using the velocity verlet algorithm [40] with a time step of 1 fs. At first, the energy of the system is minimized using the conjugate gradient (CG) minimization scheme [40]. Before applying the tensile load, NVE and NPT equilibration are carried out each for 20 ps [40]. Then, the simulation with an NVT ensemble is performed for 10 ps [40]. Finally, a uniaxial strain is applied along the loading direction at a constant strain rate of $10^9 \, s^{-1}$ under NVT ensembles to control temperature fluctuations [40]. We acknowledge that the strain rate is significantly higher than the typical experiments, however, such high rates are routinely used in MD simulations to study the materials' behavior within a reasonable simulation time and with moderate computational resources [40]–[42]. The atomic stresses in our simulations are calculated using Virial stress theorem [40], [41] which stands as,

$$\sigma_{virial}(r) = \frac{1}{\Omega} \Sigma_i (-\dot{m}_i \dot{u}_i \otimes \dot{u}_i + \frac{1}{2} \Sigma_{j \neq i} r_{ij} \otimes f_{ij})$$

In this theorem, the summation is performed over all the atoms occupying a volume, $\dot{m}_i$ presents the mass of atom i, $\otimes$ is the cross product, $\dot{u}_i$ is the time derivative, which indicates the displacement of the atom with respect to a reference position, $r_{ij}$ represents the position vector of the atom, and $f_{ij}$ is the interatomic force applied on atom i by atom j [40], [41].

To validate our computational approach and the interatomic potential parameters, we calculated the lattice constant of $Si_{0.5}Ge_{0.5}$ and Young's modulus of [100] oriented alloy with a cross-sectional area of 15 nm$^2$ at 300K and compared them with published literature [29], [31].

Table 1: Comparisons of mechanical properties between present calculation and existing literature

| Property | This study | Literature |
|---|---|---|
| Lattice parameter of $Si_{0.5}Ge_{0.5}$ alloy after NPT relaxation | 0.557 nm | 0.545 nm [29] |
| Young's modulus of $Si_{0.5}Ge_{0.5}$ alloy | 91.89 GPa | ~ 86 GPa [31] |

The excellent agreement between our simulation predictions and literature data indicates that the potential used in this study is well capable of describing the atomic interactions considered in this study. Theis observation motivate us to further investigate the mechanical properties and detailed failure behavior of the Si-Ge alloys.

## 3. Results and Discussions

**3.1 Size and temperature dependent mechanical properties of [100]-oriented $Si_{0.5}Ge_{0.5}$ alloy:** The stress-strain response of a $Si_{0.5}Ge_{0.5}$ alloy of 15nm$^2$ cross-sectional area under uniaxial tensile loading for 100 K-600 K is shown in Fig. 2(a). From the stress-strain diagram, it is clear that up to a certain point (4% strain) stress increases linearly with the increasing strain thus material exhibits elastic behavior in that region [40], [41]. After that, stress increases non-linearly and reaches its peak called ultimate tensile strength (UTS) and then decreases abruptly with a small increase in strain [40], [41]. The sharp decrease in stress indicates a brittle-type failure of the $Si_{0.5}Ge_{0.5}$ alloy. At 100 K, the ultimate strength is calculated as 18.6 GPa, with a failure strain of about 35%. As the temperature increased from 100 K to 600 K, the ultimate tensile strength of alloy decreased to 14.21 GPa, with a much lower value of failure strain of about 22.23%. This behavior can be attributed to the temperature-induced weakening of the chemical bonds in the crystal and higher thermal vibrations at elevated temperatures, as reported in earlier literature [40], [41]. To observe the effects of size on the uniaxial tensile behavior of $Si_{0.5}Ge_{0.5}$ alloy at the nanoscale, three different cross-sectional areas are chosen that are 11.05 nm$^2$, 15.00 nm$^2$, and 19.65 nm$^2$. The corresponding results at 300 K are plotted in Fig. 2(b). It is evident from this figure that the stress-strain response depends on the cross-sectional area of the alloy at the nanoscale, although material properties such as ultimate tensile strength and Young's modulus are independent of material geometry at the bulk level. Therefore, we can conclude that the analysis of continuum mechanics may not be appropriate to determine material properties at the nanoscale. The observed characteristic behaviors of nanomaterials that are different from their bulk counterparts have been reported in previous literature as well [43]–[45]. It occurs mainly due to the de-cohesion effect [46] of the surface atoms as, at this size scale, the alloy contains a large percentage of surface atoms.

These results are further translated into Fig. 2(c-d), which represent the variations of ultimate tensile strength and Young's modulus as a function of temperature for different cross-sectional areas. Young's modulus is obtained by fitting the stress-strain curve to a straight line and strain value less than 4% [40], [41]. It can be observed that there is a significant impact of size on the mechanical properties of [100] oriented $Si_{0.5}Ge_{0.5}$ alloy. Both the ultimate tensile strength and Young's modulus show an inverse relation with temperature for all cross-sectional areas. Moreover, as the cross-sectional area increases, both the ultimate tensile strength and Young's modulus of $Si_{0.5}Ge_{0.5}$ alloy increase as well. It is generally observed that the surface to volume ratio is mainly accountable for the size-dependent behavior of nanowires failure [42]. Therefore, it can be said that the surface atoms have a significant effect on the tensile properties of $Si_{0.5}Ge_{0.5}$ alloy. Besides, the smaller cross-sectional nanowires have a comparatively larger specific surface area, the lattice defects mainly occur on the surface of the nanostructure [42]. Therefore, $Si_{0.5}Ge_{0.5}$ alloy having smaller cross-sectional areas possess a higher ratio of lattice defects compared to larger cross-sectional areas [42]. The degree of lattice defects of the materials directly affects the ultimate tensile stress and Young's modulus of the $Si_{0.5}Ge_{0.5}$ alloy.

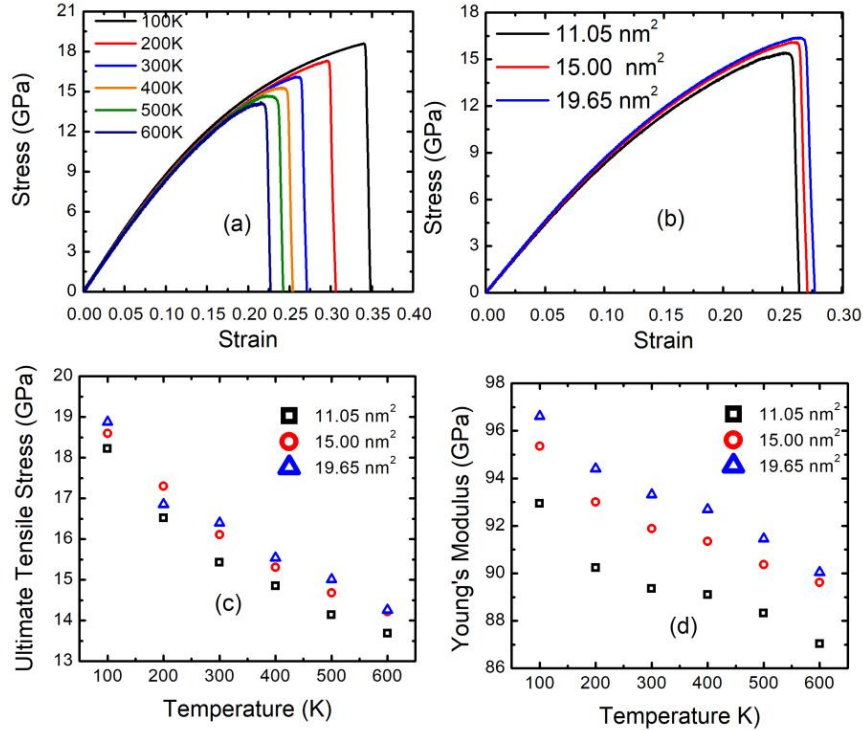

Figure 2. (a) Stress-strain curve of [100]-oriented $Si_{0.5}Ge_{0.5}$ alloy of 15 nm$^2$ cross-sectional area for different temperature (b) Stress strain curve of $Si_{0.5}Ge_{0.5}$ for different cross-sectional area at fixed 300 K (c) Variations of ultimate tensile stress for different cross-sectional areas of $Si_{0.5}Ge_{0.5}$ alloy with temperature (d) Variations of Young's modulus for different cross-sectional area of $Si_{0.5}Ge_{0.5}$ alloy with temperature.

**3.2 Effect of crystal orientation on the mechanical properties of $Si_{0.5}Ge_{0.5}$ alloy:** We investigate the effect of crystal orientation on the mechanical properties of $Si_{0.5}Ge_{0.5}$ alloy. Figure 3(a) shows the stress-strain curves of $Si_{0.5}Ge_{0.5}$ alloy at 300 K temperature with [100], [110], and [111] crystal orientations. It can be seen that [111] and [100] oriented $Si_{0.5}Ge_{0.5}$ alloys exhibit the highest and lowest ultimate tensile stress, respectively. This observation can be ascribed to the surface energy of materials since the surface energy for [111] crystal orientation is the lowest, therefore it has the highest fracture strength followed by [110] and [100] orientations [47]. Such orientation-dependent mechanical properties are also reported for Si and CdTe nanomaterials [40], [48]. The [110]-oriented $Si_{0.5}Ge_{0.5}$ alloy shows higher fracture strain compared to the other two crystal orientations. The area under the stress-strain curves indicates that it is the largest for [111] orientation. Therefore, the fracture toughness, and therefore, the total energy absorbed before the fracture is the maximum for [111] orientation and the minimum for [100] orientation. Figures 3(b-c) represent the variations of the ultimate tensile stress and Young's modulus for different crystal orientations as a function of temperature. The Young's modulus of [100], [110], and [111] oriented $Si_{0.5}Ge_{0.5}$ alloy are obtained by fitting the stress-strain curve to a straight line and strain value less than 4% as mentioned earlier [40], [41]. From Fig. 3(b), it is evident that the rate of diminution of ultimate tensile stress with temperature is the largest for [111] orientation, and for [100] orientation, it becomes the lowest. [111] crystal orientation shows the largest fracture strength among three crystal orientations then followed by [110] and [100] orientation. The [111] orientation has the highest fracture strength in the overall temperature regime, however, interestingly, in the temperature range of 100K to 200K, [110] orientation provides the highest value of Young's modulus. When a tensile load is applied in [110] direction, due to the atomic arrangements in the crystal plane, fast accumulation of stress occurs at relatively lower temperatures which leads to a higher Young's modulus compared to [111] crystallographic direction. It can be also observed that Young's modulus in [100] crystallographic direction is quite independent of the temperature.

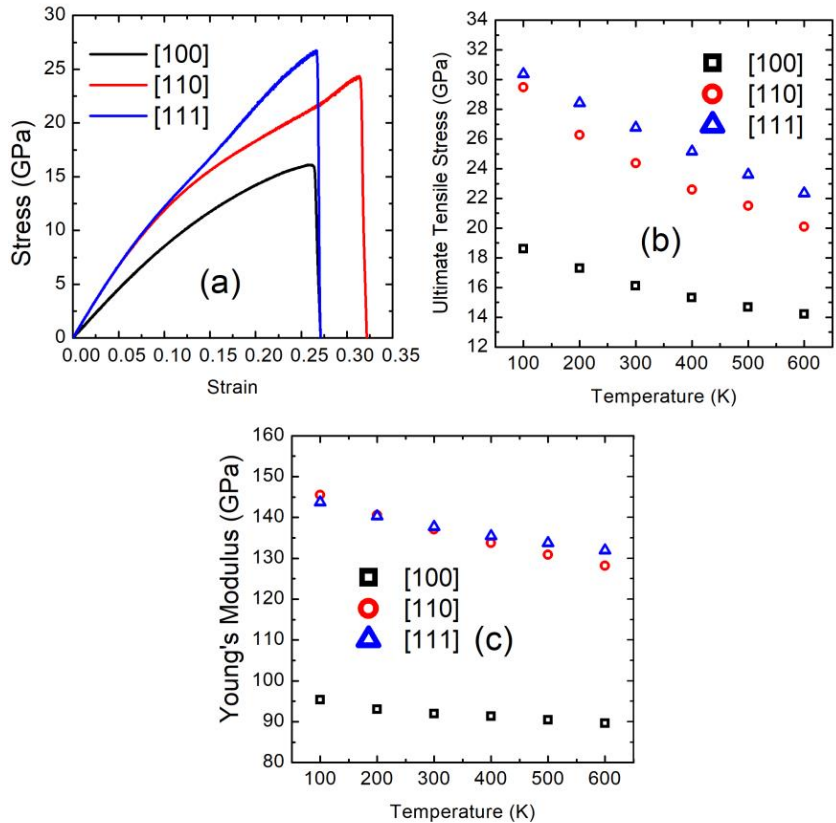

Figure 3. (a) Stress-strain curve of 15 nm$^2$ $Si_{0.5}Ge_{0.5}$ for different crystal orientation at fixed 300K (b) Variations of ultimate tensile stress for different crystal orientation of $Si_{0.5}Ge_{0.5}$ alloy with temperature (c) Variations of Young's modulus for different crystal orientation of $Si_{0.5}Ge_{0.5}$ alloy with temperature.

**3.3 Effect of vacancy defects on the mechanical properties of $Si_{0.5}Ge_{0.5}$ alloy:** According to Griffith's theory, the shape and size of defects play a more significant role than the bonding strength in determining the fracture strength [49]. Among various types of defects, the vacancy defect has pronounced effects to govern materials' physical properties [50]. More specifically, inducing only 2% of randomly distributed mono-vacancies can deteriorate the fracture strength by ∼40% in graphene [51]. In this study, we only considered the effect of randomly distributed mono-vacancies, introduced via removal of Si and Ge atoms from the alloy, on materials strength. The variation of uniaxial tensile mechanical properties of the $Si_{0.5}Ge_{0.5}$ as a function of Si and Ge vacancy concentrations is illustrated in Fig. 4 and Fig. 5, respectively. The impact of these defects arises from the under coordinated atoms and formation of dangling bonds around the random vacancies of the structure [54]. From Fig 4(a), it is evident that as the percentage of monoatomic Si-vacancy defects increases, the value of ultimate stress decreases continuously. These results are more profoundly illustrated in Fig 4(b-c). Both the ultimate tensile stress and Young's modulus decrease almost linearly with the increased concentration of vacancy defects. Introducing only 5% vacancy causes a reduction in the fracture strength and Young's modulus of about 14.21% and 11.15%, respectively. These results can be attributed to the randomly missing atoms breaking the integrity of the stress field. Bond breaking occurs when the atomic fluctuation overcomes the cohesive energy and thus, results in a disordered structure and an increment in the systems' potential energy. Removing an atom from the structure creates a hole and dangling bonds around the nearby atoms. This ultimately allows more fluctuation of the constituent bonds and results in chemical instabilities near the removed atoms. Consequently, both stress concentration and mechanical nonequilibrium arise in a stretched alloy matrix [53]. This stress concentration around the vacancy defect eventually, results in the destruction of the bond structure of the nearby members and the formation of the initial crack [50]. Therefore, the bonds at the defect region can break much earlier during the deformation process and missing bonds induces a larger deterioration of the structure, makes the structure easier to deform, and facilitate failure at a lower strain. As the percentage of defect increases, the ultimate strength and the elastic modulus of the structure substantially weakens due to the effect of stress concentration and the generation of more slip systems in the materials [54].

The effects of the removal of some random Ge atoms from the $Si_{0.5}Ge_{0.5}$ alloy are elucidated in Fig 4(d-f). The stress-strain relation, ultimate tensile strength, and Young's modulus show analogous behavior that of the removal of Si atoms. Interestingly, it can be observed that the ultimate tensile strength and Young's modulus are independent of the type of atoms removed from the alloy. This can be explained by the fact that the vacancy defects facilitate primarily the initial deformation and structural yielding. The subsequent deformations and eventual failure are related to plastic flows, and nucleation and interactions of dislocations [54].

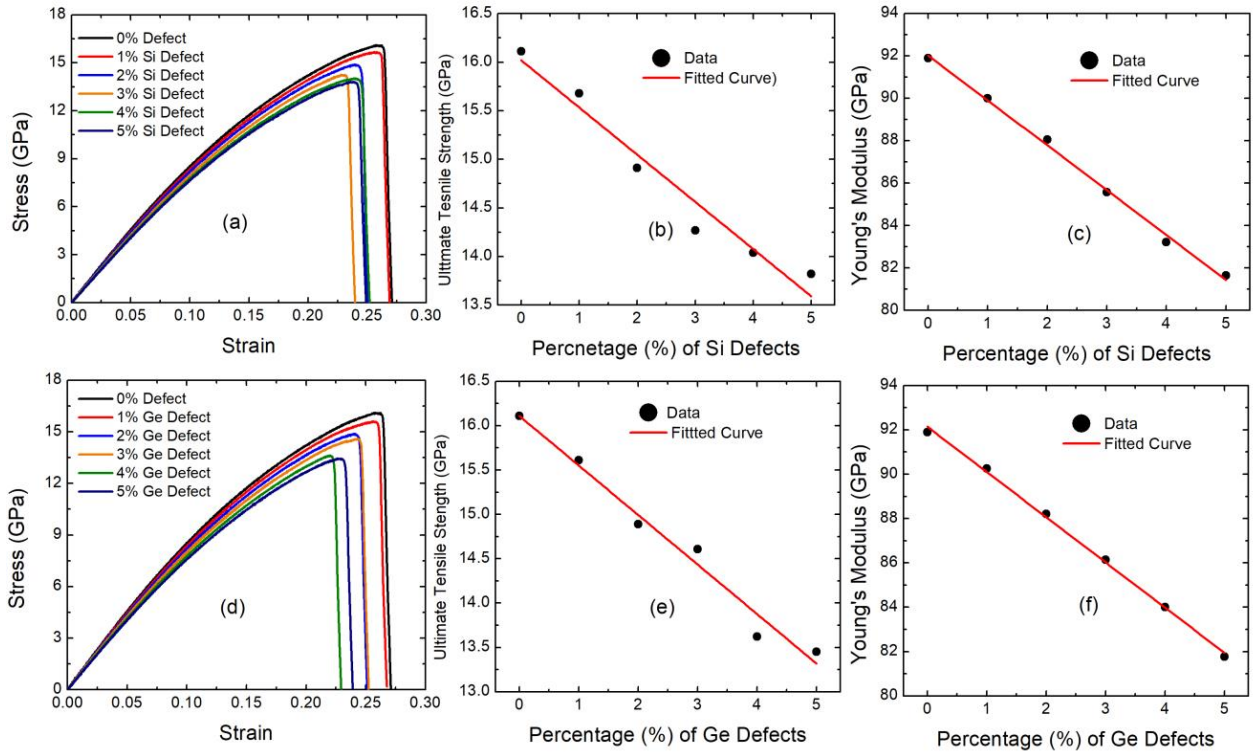

Figure 4. (a) Stress-strain curve of 15 nm$^2$ [100] oriented Si$_{0.5}$Ge$_{0.5}$ for different percentage of Si defects at 300 K (b) Variations of ultimate tensile stress of Si$_{0.5}$Ge$_{0.5}$ alloy with percentage of Si defects (c) Variations of Young's modulus of Si$_{0.5}$Ge$_{0.5}$ alloy with percentage of Si defects. (d) Stress-strain curve of 15 nm$^2$ [100] oriented Si$_{0.5}$Ge$_{0.5}$ for different percentage of Ge defects at 300 K (e) Variations of ultimate tensile stress of Si$_{0.5}$Ge$_{0.5}$ alloy with percentage of Ge defects (f) Variations of Young's modulus of Si$_{0.5}$Ge$_{0.5}$ alloy with percentage of Ge defects.

**3.4 Failure characteristic of Si$_{0.5}$Ge$_{0.5}$ alloy:** The failure behavior of Si$_{0.5}$Ge$_{0.5}$ alloy corresponding to different strain values at (a) 100 K and (b) 600 K is elucidated in Fig. 6. The process of fracture is profoundly dependent on the mechanism of crack propagation. In the case of ductile material, extensive plastic deformation typically occurs ahead of the crack tip. Ductile materials fail due to this plastic deformation. When the applied load crosses the ultimate tensile strength, the cross-section cannot sustain the applied load which causes necking and subsequently fracture of the material [55]. However, in the case of brittle materials, plastic deformation is usually absent ahead of the crack tip. In the location of cracks, stress concentration occurs, and the material instantaneously fails due to rapid crack propagation [55]. The rapid crack propagations are observed nearly perpendicular to the direction of applied stress [55]. The crack often propagates via cleavage of atomic bonds along specific crystallographic planes (cleavage planes), where shear stress becomes maximum. At 100 K temperature (Fig. 6(a)), prior to 34.18% applied strain, no structural defects appear in the alloy. At 34.18% strain, crack nucleation and primary displacement of atoms start to occur. Upon further stretching, the crack propagation continues and the effective cross-sectional area continuously reduces. When the strain reaches 34.33%, an interesting observation is found. At this point, the cross-section eventually resembles the form of a neck, like a ductile material. At 600K temperature (Fig. 6(b)), as the bond strength between Si-Si, Si-Ge, and Ge-Ge decreases, deformation in the alloy is found at 22.15% strain which is significantly earlier than 100 K temperature. The formation of the neck starts with only about 22.26% strain. At approximately 22.40% strain, it reaches at almost failure condition which is 34.84% earlier than 100 K temperature. An important observation is that, though the alloy is mainly a brittle material, it forms a neck which is the process of ductile material failure. However, note that, in ductile material, the failure process proceeds relatively slowly as the crack length extends. It requires relatively higher strain to go through the step-by-step failure processes that are necking, cavity formation, cavity coalescence to form an elliptical crack, crack propagation, and fracture [55]. However, in the case of $Si_{0.5}Ge_{0.5}$ alloy, neck formation, and failure occur very rapidly. Another difference in the failure process at the two temperatures is that, at 600 K, the stresses are concentrated at only one location but at 100 K temperature, besides necking, some secondary crack nucleation and propagation also start at some other locations prior to failure (marked by the red rectangles). This might be due to the fact that since at 600 K the failure occurs at a much lower value of strain, the system does not get enough time to initiate the secondary cracks.

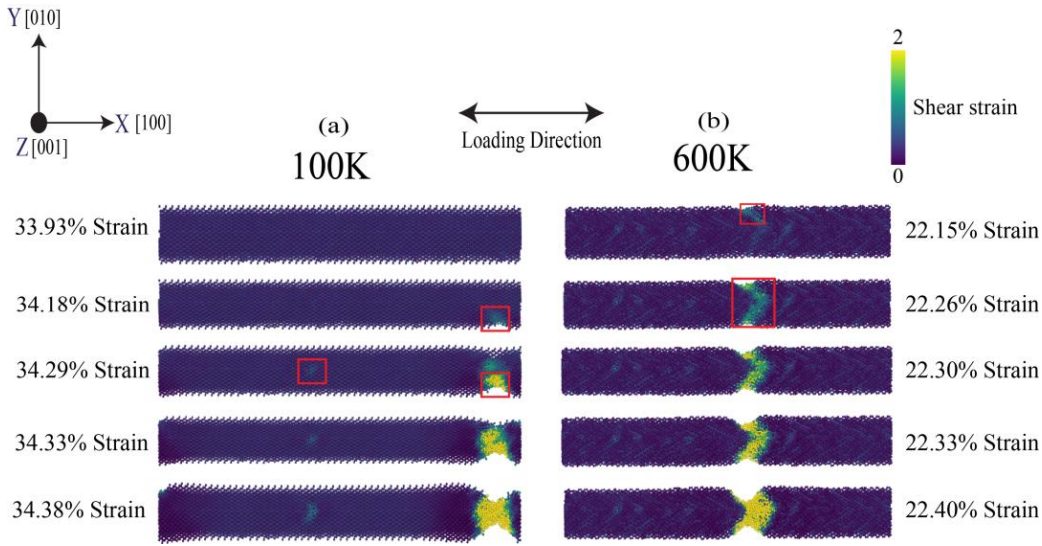

Figure. 5 Failure mechanism of $Si_{0.5}Ge_{0.5}$ at (a) 100K and (b) 600K for various shear strain level.

## 4. Conclusions

In this investigation, the mechanical properties and failure mechanism of $Si_{0.5}Ge_{0.5}$ alloy are investigated using molecular dynamics simulation. We investigated mechanical characteristics of this alloy by varying different parameters such as temperature, the cross-sectional area of the nanowire, loading direction, and finally introducing monoatomic vacancy defects by removing Si and Ge atoms randomly from the alloy matrix. Temperature plays an important role in controlling the mechanical properties of $Si_{0.5}Ge_{0.5}$ alloy. The stress-strain curve shows brittle type failure, and no brittle to ductile transition temperature is found. Both ultimate tensile strength and Young's modulus show a strong inverse relationship with temperature under tensile loading. Reduction in the cross-sectional area results in lower fracture stress and Young's modulus. The loading in different crystal orientations has a significant effect on the mechanical properties of $Si_{0.5}Ge_{0.5}$. One of the major findings of this study is that [111] direction exhibits maximum strength while the maximum fracture toughness is obtained in the case of [110] direction at 300K. The Young's modulus is higher for [110] crystal orientation for temperature 100K to 200K, but from 300K-600K, [111] orientation shows the largest Young's modulus among three crystal directions. The atomic scale defect, such as vacancy also plays a prominent role in dictating mechanical performance. Both the failure strength and the elastic modulus shows a strong inverse relationship with the increasing vacancy concentration but independent of the type of atoms removed. The failure behaviors corresponding to various strain levels in terms of shear strain parameter of $Si_{0.5}Ge_{0.5}$ alloy are also elucidated for 100 K and 600 K temperature. Upon stretching the cross-section eventually resembles the form of a neck like a ductile material. Crack nucleation and propagation start earlier in high temperatures as the chemical bonds in the Si-Ge crystal experience higher thermal fluctuation and failure occurs earlier in the case of 600K. This investigation provides a comprehensive understanding of the mechanical properties and fracture phenomena of $Si_{0.5}Ge_{0.5}$ alloy structures and will guide experimental studies and design of high-performance thermoelectric devices.

## Acknowledgement

Authors of this article would like to thank the Department of Mechanical Engineering, Bangladesh University of Engineering and Technology (BUET) and Multiscale Mechanical Modelling and Research Network (MMMRN). MMI acknowledges startup funds from Wayne State University. We sincerely acknowledge the contribution of Mr. Tawfiqur Rakib for his insightful comments on the paper and thank Mr. Rahul Jayan for helping in running simulations for this research.